\documentclass[sn-mathphys]{sn-jnl}

\usepackage{amsmath}
\usepackage{apjfonts}
\usepackage{amssymb}
\usepackage{mathrsfs}
\usepackage{natbib}
\usepackage{color}
\usepackage{etoolbox}
\usepackage{geometry}
\def\citea#1{\citep{#1}}
\def\citea#1{}

\begin{document}

\newcommand{\PrimaryMass}{\ensuremath{39_{-7}^{+6}}}
\newcommand{\PrimaryMassMedian}{39.0}
\newcommand{\PrimaryMassUpper}{45.04}
\newcommand{\PrimaryMassLower}{32.16}
\newcommand{\SecondaryMass}{\ensuremath{22_{-4}^{+8}}}
\newcommand{\SecondaryMassMedian}{22.3}
\newcommand{\SecondaryMassUpper}{29.93}
\newcommand{\SecondaryMassLower}{18.5}
\newcommand{\ChirpMass}{\ensuremath{26_{-1}^{+2}}}
\newcommand{\ChirpMassMedian}{26.0}
\newcommand{\ChirpMassUpper}{27.68}
\newcommand{\ChirpMassLower}{24.33}
\newcommand{\TotalMass}{\ensuremath{62_{-3}^{+3}}}
\newcommand{\TotalMassMedian}{62.1}
\newcommand{\TotalMassUpper}{65.34}
\newcommand{\TotalMassLower}{59.48}
\newcommand{\MassRatio}{\ensuremath{0.6_{-0.2}^{+0.4}}}
\newcommand{\MassRatioMedian}{0.6}
\newcommand{\MassRatioUpper}{0.93}
\newcommand{\MassRatioLower}{0.41}
\newcommand{\PrimarySpin}{\ensuremath{0.9_{-0.5}^{+0.1}}}
\newcommand{\PrimarySpinMedian}{0.9}
\newcommand{\PrimarySpinUpper}{0.99}
\newcommand{\PrimarySpinLower}{0.42}
\newcommand{\SecondarySpin}{\ensuremath{0.5_{-0.4}^{+0.5}}}
\newcommand{\SecondarySpinMedian}{0.5}
\newcommand{\SecondarySpinUpper}{0.95}
\newcommand{\SecondarySpinLower}{0.05}
\newcommand{\PrimaryTilt}{\ensuremath{1.4_{-0.5}^{+0.4}}}
\newcommand{\PrimaryTiltMedian}{1.4}
\newcommand{\PrimaryTiltUpper}{1.78}
\newcommand{\PrimaryTiltLower}{0.95}
\newcommand{\SecondaryTilt}{\ensuremath{2_{-1}^{+1}}}
\newcommand{\SecondaryTiltMedian}{2.0}
\newcommand{\SecondaryTiltUpper}{2.76}
\newcommand{\SecondaryTiltLower}{0.55}
\newcommand{\ChiP}{\ensuremath{0.9_{-0.4}^{+0.1}}}
\newcommand{\ChiPMedian}{0.9}
\newcommand{\ChiPUpper}{0.98}
\newcommand{\ChiPLower}{0.48}
\newcommand{\RhoP}{\ensuremath{4_{-2}^{+3}}}
\newcommand{\RhoPMedian}{4.0}
\newcommand{\RhoPUpper}{6.04}
\newcommand{\RhoPLower}{1.06}
\newcommand{\ChiEff}{\ensuremath{0.06_{-0.12}^{+0.13}}}
\newcommand{\ChiEffMedian}{0.06}
\newcommand{\ChiEffUpper}{0.18}
\newcommand{\ChiEffLower}{-0.07}
\newcommand{\Distance}{\ensuremath{1000_{-200}^{+200}}}
\newcommand{\DistanceMedian}{1000.0}
\newcommand{\DistanceUpper}{1213.34}
\newcommand{\DistanceLower}{794.52}
\newcommand{\Redshift}{\ensuremath{0.21_{-0.04}^{+0.03}}}
\newcommand{\RedshiftMedian}{0.21}
\newcommand{\RedshiftUpper}{0.24}
\newcommand{\RedshiftLower}{0.16}
\newcommand{\ThetaJN}{\ensuremath{0.5_{-0.3}^{+0.3}}}
\newcommand{\ThetaJNMedian}{0.5}
\newcommand{\ThetaJNUpper}{0.8}
\newcommand{\ThetaJNLower}{0.22}
\newcommand{\NetworkSNR}{\ensuremath{26.9_{-0.2}^{+0.2}}}
\newcommand{\NetworkSNRMedian}{26.9}
\newcommand{\NetworkSNRUpper}{27.05}
\newcommand{\NetworkSNRLower}{26.62}
\newcommand{\NumberOfSamples}{\ensuremath{1.93 \times 10^5}}
\newcommand{\LSNR}{20.95}
\newcommand{\HSNR}{14.79}
\newcommand{\VSNR}{6.65}
\newcommand{\NetworkSNRunc}{\ensuremath{26.41^{+0.2}_{-0.24}}}
\newcommand{\NetworkSNRgwtc}{26.5}
\newcommand{\LSNRgwtc}{21.2}
\newcommand{\HSNRgwtc}{14.6}
\newcommand{\VSNRgwtc}{6.3}

\title{General-relativistic precession in a black-hole binary}

\author*[1]{\fnm{Mark} \sur{Hannam}} \email{hannammd@cardiff.ac.uk} 
\equalcont{These authors contributed equally to this work}
\author[1]{\fnm{Charlie} \sur{Hoy}}
\equalcont{These authors contributed equally to this work}
\author[1]{\fnm{Jonathan E.} \sur{Thompson}}
\equalcont{These authors contributed equally to this work}
\author[1]{\fnm{Stephen} \sur{Fairhurst}}
\author[1]{\fnm{Vivien} \sur{Raymond}}

\author[2]{\fnm{Marta} \sur{Colleoni}}
\author[3]{\fnm{Derek} \sur{Davis}}
\author[2]{\fnm{H\'ector} \sur{Estell\'es}}
\author[4]{\fnm{Carl-Johan} \sur{Haster}}
\author[5]{\fnm{Adrian} \sur{Helmling-Cornell}}
\author[2]{\fnm{Sascha} \sur{Husa}}
\author[2]{\fnm{David} \sur{Keitel}}
\author[4]{\fnm{T. J. } \sur{Massinger}}
\author[6]{\fnm{Alexis} \sur{Men\'endez-V\'azquez}}
\author[7]{\fnm{Kentaro} \sur{Mogushi}}
\author[8]{\fnm{Serguei} \sur{Ossokine}}
\author[3]{\fnm{Ethan} \sur{ Payne}}
\author[9]{\fnm{Geraint} \sur{Pratten}}
\author[10,11,12]{\fnm{Isobel} \sur{Romero-Shaw}}
\author[13]{\fnm{Jam} \sur{Sadiq}}
\author[9]{\fnm{Patricia} \sur{Schmidt}}
\author[2]{\fnm{Rodrigo} \sur{Tenorio}}
\author[3]{\fnm{Richard} \sur{Udall}}
\author[14]{\fnm{John} \sur{Veitch}}
\author[14]{\fnm{Daniel} \sur{ Williams}}
\author[15]{\fnm{Anjali} \sur{Balasaheb Yelikar}}
\author[16]{\fnm{Aaron} \sur{Zimmerman}}

\affil[1]{\orgdiv{Gravity Exploration Institute}, \orgname{Cardiff University}, \orgaddress{{\street{The Parade}, \city{Cardiff}
\postcode{CF24 3AA} , \country{UK} }}}

\affil[2]{\orgdiv{Departament de F\'isica}, \orgname{Universitat de les Illes Balears},  \orgaddress{\street{Crta. Valldemossa km 7.5}, \city{E-07122 Palma}, \country{Spain} }}

\affil[3]{\orgdiv{LIGO Laboratory}, \orgname{California Institute of Technology},  \orgaddress{\city{Pasadena}, \state{California}, \country{USA} }}

\affil[4]{\orgdiv{LIGO Laboratory}, \orgname{Massachusetts Institute of Technology},  \orgaddress{\city{Cambirdge}, \state{MA}, \country{USA} }}

\affil[5]{\orgname{University of Oregon},  \orgaddress{\city{Eugene}, \state{OR} \postcode{97403}, \country{USA} }}

\affil[6]{\orgdiv{Institut de F\`isica d'Altes Energies (IFAE)}, \orgname{The Barcelona Institute of Science and Technology}, 
 \orgaddress{\city{Barcelona}, \country{Spain} }}

\affil[7]{\orgname{Missouri University of Science and Technology},  \orgaddress{\city{Rolla}, \state{MO}, \country{USA} }}

\affil[8]{\orgname{Max Planck Institute for Gravitational Physics},  \orgaddress{\city{Potsdam}, \country{Germany} }}

\affil[9]{\orgdiv{Institute for Gravitational Wave Astronomy \& School of Physics and Astronomy}, \orgname{University of Birmingham},  \orgaddress{\city{Birmingham} \postcode{B15 2TT}, \country{UK} }}

\affil[10]{\orgdiv{School of Physics and Astronomy}, \orgname{Monash University},  \orgaddress{\city{Clayton}, \state{VIC} \postcode{3800}, \country{Australia} }}

\affil[11]{\orgname{OzGrav: The ARC Centre of Excellence for Gravitational Wave Discovery},  \orgaddress{\city{Clayton}, \state{VIC} \postcode{3800}, \country{Australia} }}

\affil[12]{\orgname{Department of Applied Mathematics and Theoretical Physics}, \orgaddress{\city{Cambridge} \postcode{CB3 0WA}, \country{UK} }}

\affil[13]{\orgdiv{Instituto Galego de Fisica de Altas Enerxias}, \orgname{Universidade de Santiago de Compostela}, \orgaddress{\city{Santiago de Compostela}, \state{Galicia}, \country{Spain} }}

\affil[14]{\orgdiv{SUPA}, \orgname{University of Glasgow},  \orgaddress{\city{Glasgow} \postcode{G12 8QQ}, \country{UK} }}

\affil[15]{\orgname{Rochester Institute of Technology},  \orgaddress{ \city{Rochester}, \state{NY}, \country{USA} }}

\affil[16]{\orgname{University of Texas at Austin},  \orgaddress{\city{Austin}, \state{TX}, \country{USA} }}

\abstract{
The general-relativistic phenomenon of spin-induced orbital precession has not yet been observed in strong-field gravity.
Gravitational-wave observations of binary black holes (BBHs) are prime candidates, since we expect the astrophysical binary population 
to contain precessing binaries~\citep{Farr:2017uvj,LIGOScientific:2021psn}. Imprints of precession have been 
investigated in several 
signals~\citep{LIGOScientific:2020stg,LIGOScientific:2020ufj,LIGOScientific:2021djp},
but no definitive identification of orbital precession has been reported in any one of the 84 BBH observations to 
date~\citep{LIGOScientific:2018mvr,LIGOScientific:2020ibl,LIGOScientific:2021djp} by the Advanced LIGO and Virgo 
detectors~\citep{LIGOScientific:2014pky, VIRGO:2014yos}.
Here we report the measurement of strong-field precession in the LIGO-Virgo-Kagra (LVK) gravitational-wave signal GW200129. 
The binary's orbit precesses at a rate ten orders of magnitude faster than previous weak-field measurements from binary 
pulsars~\citep{Taylor:1982zz,Kramer:1998id,Burgay:2003jj,Breton:2008xy}.
We also find that the primary black hole is likely highly spinning. According to current binary population estimates
a GW200129-like signal is extremely unlikely, and therefore presents a direct challenge to many
current binary formation models.
}

\keywords{ black hole physics --- gravitational waves --- orbital precession }

\maketitle

Spin-induced general-relativistic orbital precession is exhibited in one of the events in the most recent LVK data 
release~\citep{LIGOScientific:2021djp}, GW200129\_065458, which we refer to throughout this paper as GW200129. 
We find that, given our
astrophysical priors, observational biases, and an assumption of Gaussian noise, the precessing hypothesis is favoured to the non-precessing hypothesis by a factor of at least 30:1. 
Considering noise effects alone, there is only a 1 in 25,000 chance that the imprint of precession on this signal
is entirely due to noise. 
Regarding priors, we use the most agnostic statistical priors available, i.e., flat in the component masses and spin magnitude and the 
cosine of the spin misalignment angle, and these are the same as those used in all LVK detections~\cite{LIGOScientific:2021djp}.
We also verify that the main features of our results are (a) not explained by noise, 
by repeating our analysis on a theoretical signal injected into detector data, and on a theoretical signal as it would appear in a 
noise-free detector network and (b) the most accurate and reliable available, by comparing our model to the raw output from solving Einstein's equations directly.

GW200129 was reported with a signal-to-noise ratio (SNR) of \NetworkSNRgwtc\, across the three-detector LIGO-Virgo 
network~\citep{LIGOScientific:2021djp}.
This makes it the loudest BBH signal yet observed, slightly louder than the first detection, GW150914, initially reported with a network
SNR of 24~\citep{LIGOScientific:2016aoc}. 
The LVK source properties for GW200129 were based on two analyses. One reported high precession broadly consistent with the results we
present here, and the other did not. The LVK analysis gave equal weight to both analyses, leaving the 
properties of GW200129 unclear~\citep{LIGOScientific:2021djp}.  Our work
shows that both analyses rely on theoretical gravitational waveform models that may not be sufficiently accurate to measure
GW200129. However, a third theoretical model \emph{is} sufficiently accurate and with that model we are
able to identify that GW200129 is indeed highly precessing. 
Our measurements of the binary's properties, with 90\% credible intervals, are given in Tab.~\ref{tab:parameters}.
The details of our parameter-estimation and systematics analysis are given in the Methods section.

\begin{table}
\begin{center}
\begin{tabular}{l | c l }
\hline
\hline
Primary mass, $m_1$ ($M_\odot$) & \PrimaryMass \\
Secondary mass, $m_2$ ($M_\odot$) & \SecondaryMass \\
Total mass, $M = m_1 + m_2$ ($M_\odot$) & \TotalMass \\
Mass ratio, $q = m_2/m_1$ & \MassRatio \\
Primary spin, $a_1/m_1$ & \PrimarySpin \\
Primary spin tilt angle, $\theta_{\rm LS_1}$ (rad) & \PrimaryTilt \\
Secondary spin, $a_2/m_2$ & (undetermined)\\
Binary inclination, $\theta_{\rm JN}$ (rad)& \ThetaJN \\
Luminosity distance, $D_L$ (Mpc) & \Distance \\
Redshift, $z$ & \Redshift \\
\hline
\hline
\end{tabular}
\caption{{\bf Properties of the binary-black-hole source of GW200129.} Source parameter measurements from our analysis of 
GW200129, with uncertainties at the 90\% credible interval. The posterior distributions for both the secondary spin magnitude 
and tilt angle $\theta_{LS_2}$ are essentially flat, and for this reason we have indicated the secondary spin as ``undetermined''. 
This is expected for a signal of GW200129's strength, as explained in the main text. A naive application of our 90\% credible-interval 
calculation gives $a_2/m_2 \in [0.05,0.95]$ and $\cos(\theta_{LS_2}) \in [-0.93,0.85]$, i.e., an arbitrary 90\% 
stretch of the range of possible values.
}
\label{tab:parameters}
\end{center}
\end{table}

The key feature of GW200129 that we focus on in our analysis is the measurement of strong-field general-relativistic orbital and
spin precession. Precession was previously measured in binary pulsars, both from the precession of one of the pulsars'
spins~\citep{Kramer:1998id,Breton:2008xy,Manchester:2010dh,Fonseca:2014qla,Desvignes:2019uxs}, or precession of the binary's
orbital plane~\cite{Krishnan:2020txo}. In the limit of a point particle orbiting a much larger body, the precession of the test body's spin
due to the presence of the larger body is known as de Sitter precession, while the precession of a test body's orbit due to the spin 
of a larger body is Lense-Thirring precession. In the general case, where the two bodies can be of comparable mass, the two 
effects can be seen to counteract each other, such that the direction of the total angular momentum of the binary remains 
constant and the total spin precesses at the same rate as the orbital plane~\citep{Barker:1975ae,Apostolatos:1994mx,Kidder:1995zr}. 
Black-hole mergers are in the strong-field nonlinear regime
of general relativity well beyond leading-order effects, but one quantity that can be compared across all regimes is the system's 
precession frequency. 

\begin{figure}
\begin{center}
  \includegraphics{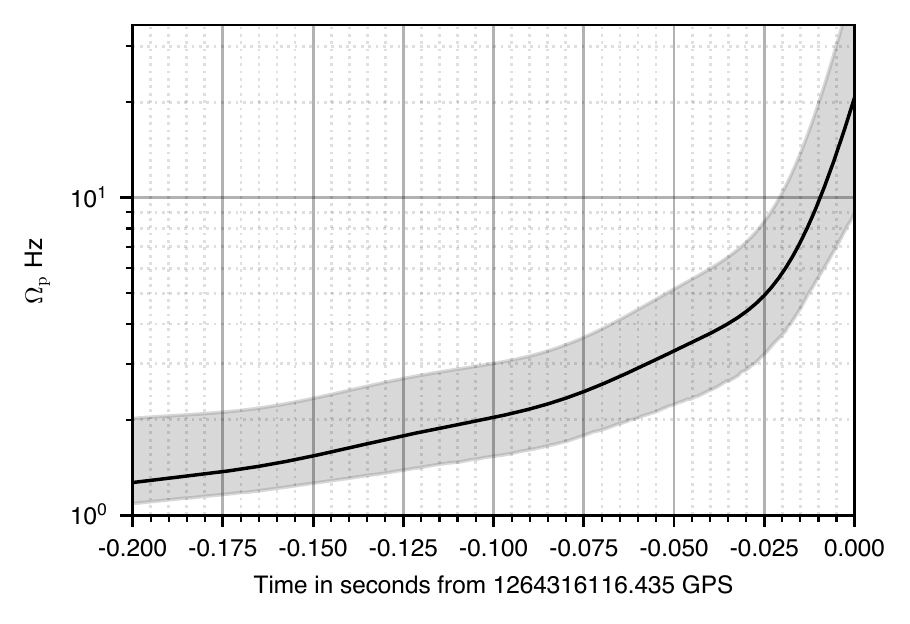}
  \caption{
  \label{fig:omega}
{\bf Precession frequency of GW200129.} The precession frequency $\Omega_{\rm p}$ of GW200129 as a function of time, from 0.2s before merger, along with the 90\% credible interval.
  }
\end{center}
\end{figure}

For example, the pulsar PSR B1913+16 was found to have a precession rate of approximately 1.2 
degrees per year~\citep{Kramer:1998id}, or $1.1 \times 10^{-10}$\,Hz. Later, the double pulsar PSR J0737-3039 was found to 
have a precession rate of about 4.8 degrees per year~\citep{Breton:2008xy}, or $4.4 \times 10^{-10}$\,Hz. In contrast, we find GW200129 
to have an average precession rate in the LIGO-Virgo band of $\sim$3\,Hz, i.e., \emph{ten orders of magnitude higher than previously 
measured.} Fig.~\ref{fig:omega} shows the precession frequency 
$\Omega_{\rm p}$ as a function of time. We see that the precession frequency is at least 1\,Hz when the signal enters the 
detector sensitivity band, and rises to over 10\,Hz at merger, which is indicated here by $t=0$, the time of maximum signal amplitude. 
The binary's orbital plane is inclined to the total angular momentum by $\sim$0.5\,rad; although this angle increases through 
inspiral~\cite{Apostolatos:1994mx}, it does not change significantly over the short duration of this observation. 
We discuss further the subtleties of this measurement in the Methods section. 

\begin{figure}
\begin{center}
  \includegraphics{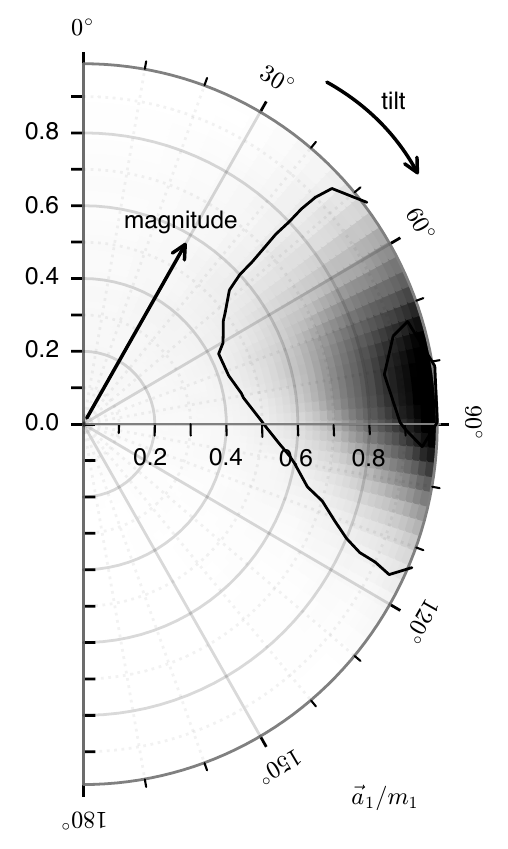}
  \caption{
  \label{fig:a1}
{\bf The magnitude and tilt of the spin of the larger black hole.} Two-dimensional posterior probability for our measurement of the dimensionless spin  $ \vec{a}_1 /m_1$ of the primary black hole, 
and its mis-alignment (tilt) with the direction of the orbital angular momentum. Tilt angles of $90^{\circ}$ means that the spin vector 
lies within the plane of the binary. The colour indicates the posterior probability
per pixel. This plot is produced by using histogram bins that are constructed
linearly in spin magnitude and the cosine of the tilt angles such that each bin
contains identical prior probability. The probabilities are marginalized over the azimuthal angles. Contours represent the 50\% and 90\% credible intervals. 
  }
\end{center}
\end{figure}

We now turn to the black hole's spin angular momentum, $S$. The spin of a black hole with mass $m$ is usually represented in 
geometric units by the dimensionless quantity $a/m = S/m^2$, which ranges from zero (non-spinning) to one (extremal spin). 
Fig.~\ref{fig:a1} shows
the posterior distribution of our measurement of the spin of the primary black hole $a_1/m_1$, and its angle of misalignment (``tilt'')
with respect to the orbital angular momentum. The misalignment is close to 90$^\circ$, and therefore the spin lies almost entirely 
in the orbital plane; this is the cause of the significant orbital precession. We also find that the primary black hole's spin is larger than 
$\sim$0.4, with a strong preference for much higher values, with $a_1/m_1 = \PrimarySpin$. 
The signal is not strong enough for 
the secondary spin to be measured; this is the case in all BBH observations to date, and in general reliable measurements of 
both spins are not expected for SNRs less than $\sim$100~\citep{Purrer:2015nkh}.

\begin{figure}
\begin{center}
  \includegraphics{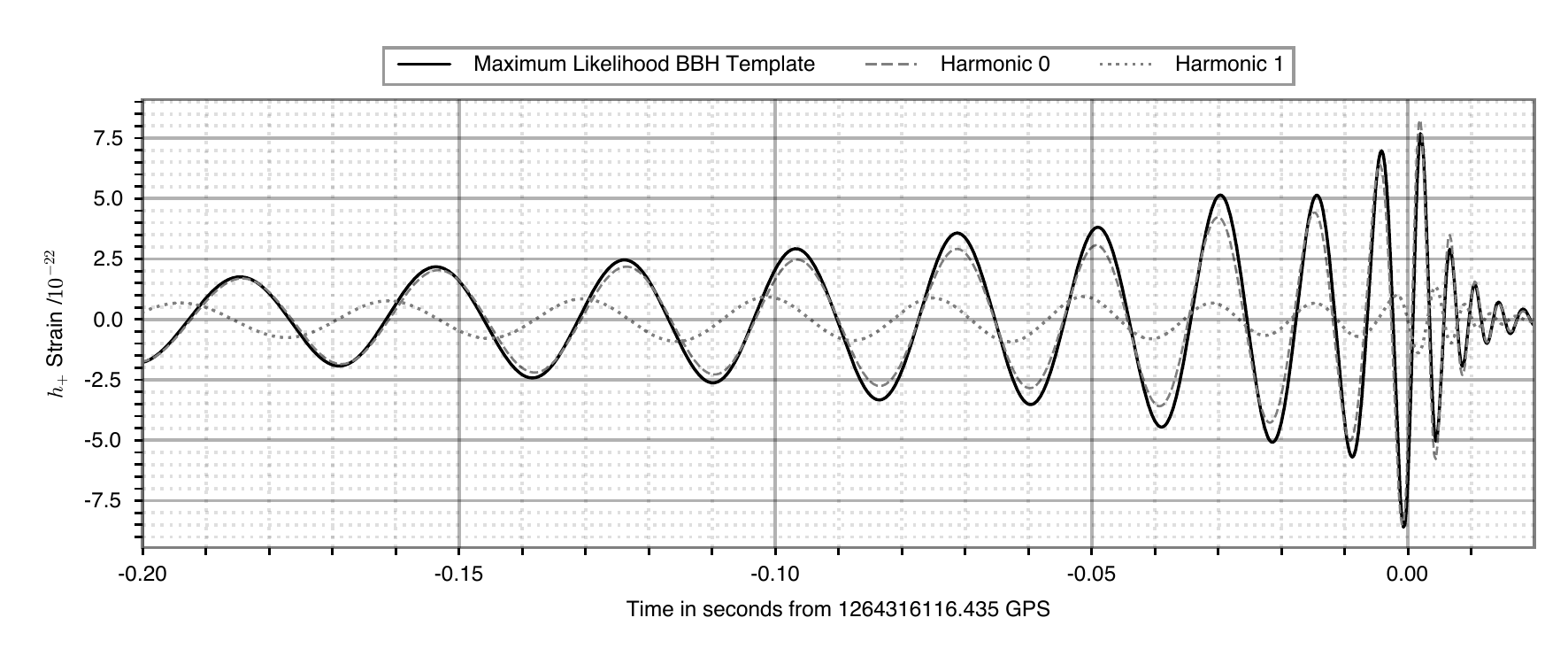}
  \caption{
  \label{fig:data}
{\bf Anatomy of GW200129.} The figure shows the theoretical maximum likelihood waveform from our 
parameter-estimation analysis (see Methods section), and its approximate decomposition into the two strongest (out of five)
 non-precessing harmonics~\citep{Fairhurst:2019vut}. It is the second harmonic that provides the power in precession. 
  }
\end{center}
\end{figure}

Black-hole spin and orbital precession leave only subtle imprints on the waveform. This is why a high SNR is in general necessary for
precession to be measured. Fig.~\ref{fig:data} shows the plus polarisation of the theoretical signal preferred by our analysis, 
over a $\sim$0.25\,s window that corresponds to roughly half of the duration of data that we analyse. 
It is possible to make an approximate decomposition of the signal into two non-precessing harmonics, and to recover
the mild precession modulations as the beating between these two harmonics~\citep{Fairhurst:2019vut}. The leading harmonic
(``harmonic 0'') makes up the dominant part of the signal, while the next harmonic (``harmonic 1'') provides the power in precession.
Fig.~\ref{fig:data} also shows these two harmonics. The harmonics are in phase at merger and roughly ninety degrees out of phase 
0.2\,s earlier, illustrating that the system undergoes roughly one precession cycle while in the detector's sensitivity band from 0.65\,s
before meger.

Having argued that the signal GW200129 was likely produced by a precessing high-spin binary, we address the question of how
it may have formed. A single observation is not informative about the overall binary population. However, GW200129 does tell us 
that black holes can form with high spins, and can end up in binaries with a large spin misalignment. The 84 LVK BBH 
observations suggest that black-hole spins are typically low, with half of spin magnitudes less than 0.26~\citep{LIGOScientific:2021psn}, 
and this is consistent with the expectation that black holes that form through stellar collapse will not be highly spinning~\citep{Fuller:2019sxi}. 

Although formation of a binary like GW200129 would be rare from the formation mechanisms that are consistent with the currently 
observed astrophysical population, there are a number of viable routes. One is a hierarchical merger, where successive black-hole mergers
could ultimately lead to a binary where one component is highly spinning, and misaligned with the secondary. Hierarchical mergers may have already been identified through gravitational wave observations; see, e.g., Refs.~\cite{Kimball:2020qyd, Tiwari:2020otp}. Another is a field binary where the binary black hole is formed from isolated stellar progenitors. Although various aspects of this field binary tend to suppress spin misalignment~\cite{Kalogera:1999tq}, there are regions of the parameter space that can produce binaries with high spins and significant precession~\cite{Steinle:2020xej}. Other options to produce high spins include chemically homogeneous evolution~\citep{Mandel:2015qlu} (although this would \emph{not} lead
to spin misalignment), formation in AGN~\citep{Tagawa:2020dxe} (including accretion),
and a binary that has formed from a triple-black-hole system~\citep{Antonini:2017tgo}. 

In some of these scenarios an event such as GW200129 would be extremely rare, and if that is the 
case then we do not expect to observe another signal of this type for some time. Indeed, following the techniques described in Ref.~\cite{Fairhurst:2019srr} and 
assuming the best estimate for the spin distribution of black holes based on current observations~\cite{LIGOScientific:2021psn}, we calculate that a GW200129-like 
system will be observed only once in every 1200 GW observations. This means that a similar system is unlikely to be observed again in the next two LVK observing runs~\cite{LIGOObservingPlans:2019aaa} and therefore GW200129 will likely be the only event with significant precession and high spin in the first decade of GW astronomy. Given the exceptional nature of 
GW200129 within current binary population estimates~\cite{LIGOScientific:2021psn}, it is more likely that GW200129 presents a significant challenge to 
the two favoured BBH formation mechanisms: field binaries~\cite{Kalogera:1999tq}, and dynamic binaries (formed when two black holes become gravitationally bound in dense stellar environments)~\cite{rodriguez2016illuminating}. In fact, we note that in the majority
of observed BBH signals the SNR has not been high enough for precession to be measured, even if the black-hole spins
are large and mis-aligned, and so these may be more common than previously thought. If so, more binaries such as this will be
observed during the upcoming LVK observing runs and they will provide clues as to the specific mechanism that produces 
high-spin, high-misalignment binaries, and how frequently they form.

\section{Methods}
\label{sec:method} 

\renewcommand{\figurename}{Extended Figure}
\setcounter{figure}{0}

We analyse public LVK data from the second half of the third observing run (O3b)~\citep{LIGOScientific:2021djp,LIGOScientific:2019lzm}, 
which ran from November 1, 2019 until March 27, 2020. The signal GW200129 was observed on January 29, 2020, and enters the
detector sensitivity band at 20\,Hz approximately 0.65s before merger. During that time the binary completes roughly nine orbits.
GW200129 was reported by the LVK with SNRs in each of the three operational detectors of \LSNRgwtc\, in Livingston, 
\HSNRgwtc\, in Hanford and \VSNRgwtc\, in Virgo. The total network SNR is \NetworkSNRgwtc.

We determine the properties of the signal source by comparing the data against theoretical predictions from general relativity. 
We calculated predicted signals using the most accurate theoretical waveform model available for configurations consistent
with this signal, {\tt NRSur7dq4}~\citep{Varma:2019csw}: when we analyse the data starting at 20\,Hz, the model is appropriate for sources 
at a redshift of 0.2 with total binary masses above 56.7\,$M_\odot$ and where the primary
black hole is no more than four times more massive than the secondary. We find
that the binary's mass is above 58\,$M_\odot$ with 99\% confidence, and the mass ratio is less than 1:3 with 99\% confidence. 

We perform our analysis using the Bayesian Markov-Chain Monte-Carlo (MCMC) code {\tt LALInference}~\citep{Veitch:2014wba} 
that has been used in LVK analyses since
the first GW detection in 2015~\citep{LIGOScientific:2016vlm}, and has been rigorously tested and refined since that time. 
We used the same prior, sampler settings, power spectral densities, and calibration envelopes as those used in LVK GWTC-3 analyses,
except for a cut in the prior parameter space to accommodate the {\tt NRSur7dq4} model, i.e., limiting the mass ratio to be less than 1:4, the 
chirp mass to be between $14.5\,M_{\odot}$ and $49\,M_{\odot}$ and the total mass to be above $68\,M_\odot$ in the detector frame, 
which corresponds to $56.7\,M_{\odot}$ at redshift 0.2. With respect to the component spins, our prior was flat in spin magnitude and the 
cosine of the misalignment angle, and flat in the azimuthal angle of each spin.

Our analysis is restricted to the $\ell \leq 3$ multipoles of the signal. The signal power in the $\ell > 3$ multipoles has 
an SNR of less than 1.0, and does not significantly change the results. There is also some power lost in the $(3,\pm3)$ multipoles
between 20\,Hz and 30\,Hz for the lowest-mass configurations that are sampled, due to these multipoles starting at a frequency $3/2$ 
times higher than the dominant $(2,\pm2)$ multipoles. We have also verified that this missing power is negligible and also does
not affect the results. Finally, we performed analyses on both the raw public detector data and the public data that had undergone glitch 
removal. The ``de-glitched'' data were used in the official LVK results, and we do the same here. The raw data gave very similar results,
with slightly increased support for precession. 

We find that the signal power due to precession has an SNR of $\rho_p = \RhoP$. If precession effects were due to noise, we expect 
$\rho_p$ to be no higher than around 2 \citep{Green:2020ptm}. 
The $\rho_p$ estimate of power in precession makes use of an approximate
two-harmonic decomposition of the signal~\citep{Fairhurst:2019vut}, which becomes less applicable for high-mass signals; 
however, in these cases it is most likely an \emph{underestimate} of the total power due to precession. 
We also find, following Ref.~\citep{Hoy:2021dqg}, that there is only a 1 in 25,000 chance that noise alone would produce the inferred
precession SNR that we measure.

Despite the clear measurement of a high in-plane primary spin in Tab.~\ref{tab:parameters} and Fig.~\ref{fig:a1}, we now quantify our 
statistical confidence that GW200129 was produced by a precessing binary. We compare the marginal likelihood from our 
parameter-estimation analysis with that from a second analysis, which restricts the in-plane spin components to be zero, i.e., 
a non-precessing binary. We calculate that the precessing-binary hypothesis is favoured over the non-precessing hypothesis
by a factor of 30:1, or a $\log_e$ Bayes factor of 3.4 ($\log_{10}$ Bayes factor of 1.5). 
This Bayes factor is consis- tent with the output from independent nested samplers~\cite{Veitch:2014wba,Smith:2019ucc, Speagle:2019ivv}. 

We estimate the time evolution of the precession frequency, and the opening angle between the orbital and total angular momenta
(i.e., the inclination of the orbital plane), from a subset of theoretical waveforms within the 90\% credible region of our parameter-estimation
results. It is possible to estimate these dynamical properties from the multipole structure of the waveforms~\citep{Schmidt:2010it}. The
precession frequency and opening angle calculated from the signal are not identical to those in the binary dynamics~\citep{Hamilton:2021pkf},
but the differences are smaller than the overall uncertainty in our measurements. The signal-based precession frequency as measured from 
the theoretical model also contains oscillations that we know are not present in the orbital dynamics (by comparison with numerical-relativity
results), and we filter those out of the results shown in Fig.~\ref{fig:omega}. 

There are four potential sources of error in our results. (1) uncertainties in the waveform model have biassed the results, 
(2) the parameter-estimation code has settled on the wrong source-parameter values, (3) prior effects, and (4) the results are due to noise 
artifacts. We now discuss each of these in detail. 

(1) Waveform model uncertainties. 
The waveform model was tuned to numerical-relativity simulations between mass ratios of 1:1 and 1:4, and black-hole spins
up to 0.8, and is well extrapolated up to extreme spins~\citep{Varma:2019csw, Islam:2020reh}.
All tests reported in the literature, and our own, suggest that the extrapolation to extreme spins is well-behaved, i.e., the 
properties of the waveforms (changes in phasing, amplitude and precession angles) extend smoothly from zero spin to extreme spin. 

The usual measure of the accuracy of a model waveform is the mismatch between the model and a fiducial ``true'' waveform. 
The mismatch between two waveforms is calculated from a noise-weighted inner product 
between the normalised waveforms; the mismatch is the deviation of this quantity from unity, and a value of 0 indicates that the waveforms
agree perfectly (up to an overall amplitude rescaling), and a value of 1 that they are completely orthogonal. The mismatch also allows
us to estimate the SNR at which two signals would be indistinguishable~\citep{Baird:2012cu}. If the mismatch between two signals is
$\mathcal{M}$, then they will be indistinguishable when,

\begin{equation}
\mathcal{M} \leq \frac{\chi_k^2(1-p)}{2\rho^2},
\end{equation}
where $\chi_k^2(1-p)$ is calculated from the cumulative distribution function of the $\chi^2$ distribution with $k$ 
degrees of freedom at probability $p$. A binary black hole system undergoing non-eccentric inspiral has eight physical parameters
(the two masses, and the components of each spin vector); with eight degrees of freedom, two waveforms will be indistinguishable 
at 90\% confidence if their mismatch satisfies $\mathcal{M} \leq 6.68/\rho^2$. Given that we do not measure most of the spin 
components, we might instead consider four degrees of freedom (e.g., the two masses, and the in-plane and aligned spin contributions),
which provides a stronger mismatch criterion of $\mathcal{M} \leq 3.89/\rho^2$.
For an SNR of 27, two signals will be indistinguishable if their mismatch is less than $9.2 \times 10^{-3}$ if we assume eight degrees of 
freedom. If we apply the more stringent criterion of four degrees of freedom, the mismatch must be below $5.3 \times 10^{-3}$ for them to 
be indistinguishable.

Within the model's calibration parameter space, the mismatch error between the model and fully general relativistic numerical relativity
simulations was less than $4 \times 10^{-3}$ for 95\% of configurations considered in Ref.~\citep{Varma:2019csw}. This is well within
both criteria proposed above, and indicates that the model is well within the accuracy requirements to measure GW200129.

\begin{figure}
\begin{center}
  \includegraphics{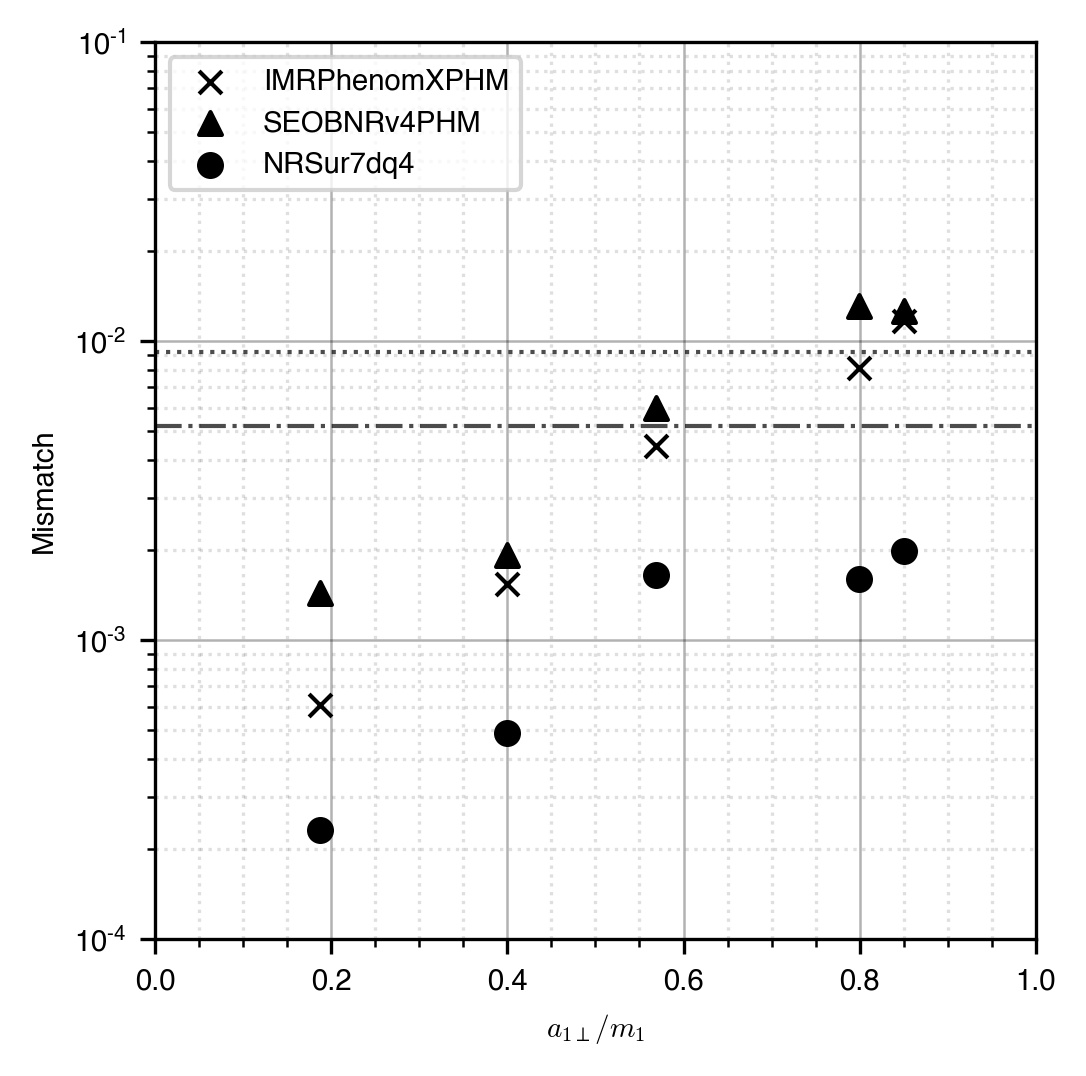}
  \caption{
  \label{fig:mismatches}
{\bf Mismatches between waveform models and numerical-relativity waveforms.} Mismatches between three waveform models, and five waveforms from numerical-relativity simulations. The simulations were of binaries
with mass-ratio 1:2, and varying values of the in-plane spin magnitude on the primary black hole, $a_{1\perp}/m_1$, which is what drives precession.
The dotted line shows the accuracy threshold assuming eight degrees of freedom, and the dashed-dotted line the threshold assuming only 
four degrees of freedom (see text for discussion). Only the {\tt NRSur7dq4} is well within the accuracy thresholds for GW200129.
  }
\end{center}
\end{figure}

We explicitly test these accuracy statements for signals in the region of parameter space of GW200129, for  {\tt NRSur7dq4} and for
the two alternative theoretical models that were used in the LVK analysis. 
These were selected because the total mass and mass-ratio limitations of the {\tt NRSur7dq4} 
model mean that it cannot be used to measure the properties of binaries with low total mass. Since most of the LVK detections to date were 
indeed at masses below the model's low-mass limit, the LVK analysis of the 
O3b data made use of two other independent models, {\tt PhenomXPHM}~\citep{Pratten:2020fqn,Garcia-Quiros:2020qpx,Pratten:2020ceb} 
and {\tt SEOBNRv4PHM}~\citep{Ossokine:2020kjp}. Both models are on average less accurate than
{\tt NRSur7dq4} over its calibration region, but have the advantage of being applicable at both higher mass ratios, and lower masses, and
their accuracy is sufficient at the SNRs and configurations (in particular, low spins) of most LVK observations. However, they may not meet
the accuracy requirements of the high-SNR high-spin merger GW200129. Extended Fig.~\ref{fig:mismatches} shows mismatches for the three different 
models against
a set of publicly available numerical-relativity waveforms~\citep{SXS:catalog}. All five simulations are at mass-ratio 1:2, and have 
different magnitudes of in-plane spin on the larger black hole.
The simluations used are SXS:BBH:1128, SXS:BBH:1096, SXS:BBH:0800, SXS:BBH:1097, SXS:BBH:1215, listed in order of 
increasing in-plane spin. 
The mismatches are calculated at a binary inclination of $\pi/6$ (close to that recovered for
GW200129), a total mass of 73\,$M_\odot$ (the maximum likelihood total mass in the detector frame), and are optimised over the template 
phase, polarization and in-plane spin direction, while SNR-weighted averaging over signal polarization and phase. (More details on the
mismatch calculation that we use are given in Ref.~\citep{Hamilton:2021pkf}.)
The figure also shows the mismatch accuracy threshold for a signal with SNR 27 with eight and four degrees of freedom. We see
that {\tt NRSur7dq4} meets the eight-degrees-of-freedom requirement for high spins, and indeed is the only model that also meets 
the requirement with four degrees of freedom. For the simulation with the highest in-plane spin of $a_{1\perp}/m_1 = 0.85$, 
SXS:BBH:1215, we find a mismatch of $2.23\times10^{-3}$, corrsponding to a distinguisable SNR of $\sim$42 (with four degrees of freedom). 
We note that this configuration is \emph{outside} the calibration region of the model, since $a_1/m_1 > 0.8$. 

\begin{figure}
\begin{center}
  \includegraphics{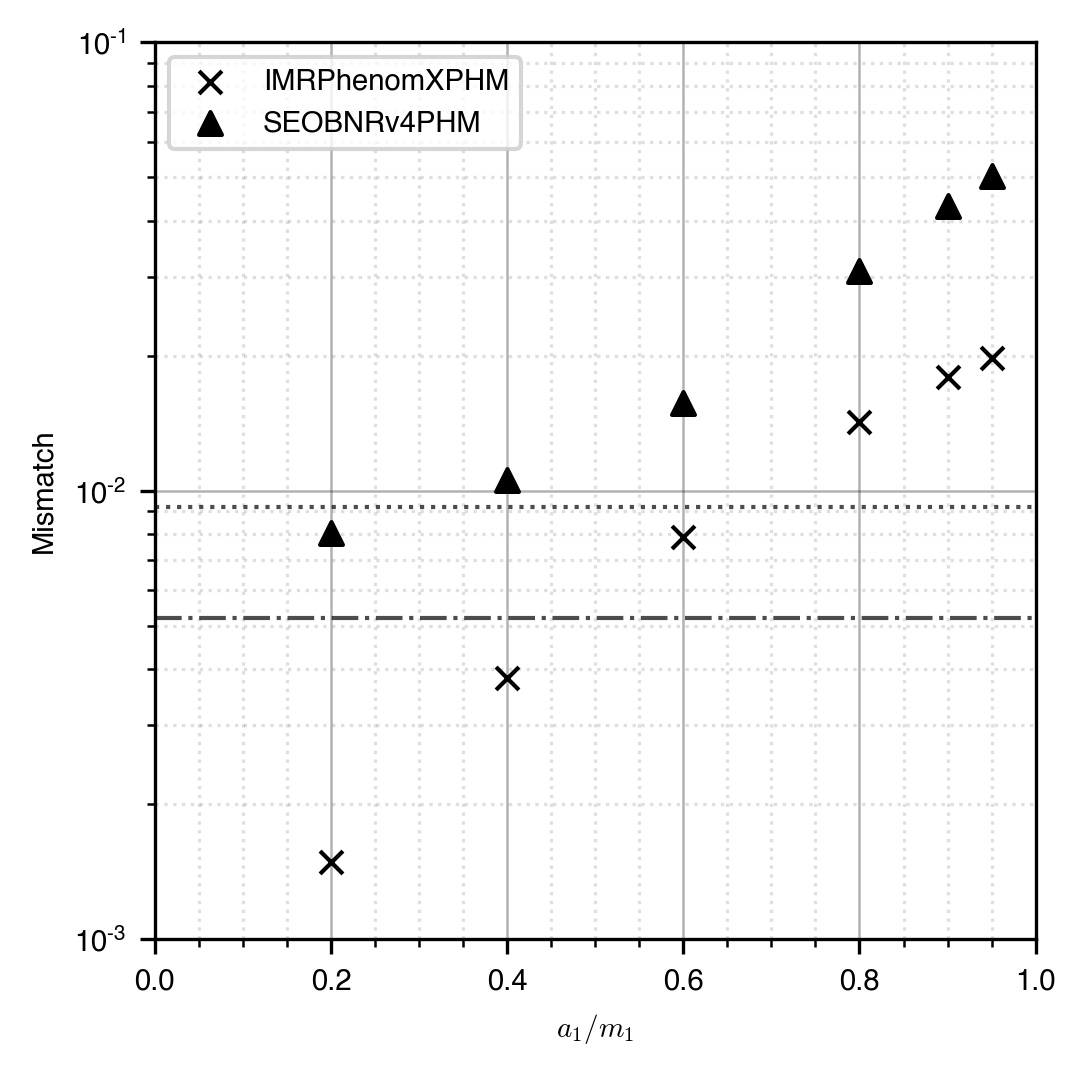}
  \caption{
  \label{fig:mismatches2}
{\bf Mismatches between the most accurate waveform model and those used in the LVK analysis.} Mismatches between theoretical signals (calculated using the {\tt NRSur7dq4} model) against the {\tt PhenomXPHM} and 
{\tt SEOBNRv4PHM} models. The model parameters are those recovered from our analysis of GW200129, but with a 
range of values of the primary spin magnitude $a_1/m_1$. We see that neither model meets the accuracy thresholds for GW200129 at high
spins, but the better agreement of {\tt PhenomXPHM} is consistent with it recovering results closer to those reported in this work.
  }
\end{center}
\end{figure}

In Extended Fig.~\ref{fig:mismatches2} we show an alternative mismatch-accuracy check. Here we assume that the {\tt NRSur7dq4} model 
accurately represents binary signals, and calculate the mismatch against that model evaluated at the parameters that yield the maximum
likelihood in our parameter-estimation analysis, and instances of the {\tt PhenomXPHM} and {\tt SEOBNRv4PHM} models evaluated
at the same parameters, and optimised as in Extended Fig.~\ref{fig:mismatches}. We also calculate mismatches for model evaluations with
the same parameters, but with a range of values of the primary spin magnitude $a_1/m_1$. We see again that both models do not
meet the indistinguishability requirement when the spins are high. We also see that the high-spin mismatches are better for the 
{\tt PhenomXPHM} model, which is consistent with that model recovering signs of precession in GW200129, while 
{\tt SEOBNRv4PHM} does not. Note, however, that systematics errors are likely still a problem for {\tt PhenomXPHM}, as
discussed in Ref.~\citep{Biscoveanu:2021nvg}. 

A final source of systematic uncertainty is that we assume non-eccentric inspiral. As an example of how this can affect a measurement,
the earlier GW observation GW190521 also appears to have a large in-plane spin when analysed with a non-eccentric-binary 
model~\cite{LIGOScientific:2020ufj}, but the apparent precession signature may be due to orbital 
eccentricity~\citep{Romero-Shaw:2020thy,Gayathri:2020coq} or even a head-on collision~\citep{CalderonBustillo:2020odh}. 
The key difference with GW200129 is that the binary has lower mass, and so more GW cycles are detectable, and over this number of cycles 
it should be possible to distinguish eccentricity from precession; GW200129 undergoes roughly one precession cycle in the detectors'
sensitivity bands, while eccentricity 
appears at the orbital timescale. Therefore orbital eccentricity would produce $\sim$10 modulations in the amplitude and phase of the 
observed signal, while precession would only produce one modulation. 

(2) Parameter-estimation uncertainties. 
The results have several features that at first sight raise concerns. 
In preliminary parameter-estimation runs the one-dimensional posterior distribution functions for the masses were bi-modal, which is 
often a sign of a noise artefact, or that the MCMC method has not converged, or some other issue. However, we found that the results
became much cleaner when the sampler was run for longer, and the final production run produced \NumberOfSamples \, samples. 
(By comparison with standard GW applications of the {\tt LALinference} sampler, we would normally consider $\sim$$10^4$ samples 
to be sufficient.) In the final results, the probability distribution for the binary's mass ratio has a tail that
extends to equal masses. However, our astrophysical priors have a preference for equal masses, and so this result is not surprising,
and, despite this prior preference, over 80\% of the samples are at mass ratios above 1:1.35.

Another concern is the apparent ``railing'' of the spin measurement against extremal spin;
this can also be a sign of noise issues. There have also been studies that suggest that the astrophysical prior and observational biases
will pull the spin-magnitude measurement down to lower values, even if the source contains a highly spinning black 
hole~\cite{Chatziioannou:2018wqx}. However, in those studies the large spin was aligned with the orbital angular momentum, and the 
prior on the aligned-spin components has a strong preference for low spin. In our case, the priors on the spin magnitudes are flat. 

Finally, we find that if we restrict the analysis to the Livingston and Virgo detectors, then a similar precessing configuration is 
recovered, but if we restrict to Hanford and Virgo, we recover a configuration closer to equal-mass, and with minimal precession.
This suggests that the data in the Livingston and Hanford detectors may not be consistent. 

\begin{figure}
\begin{center}
  \includegraphics{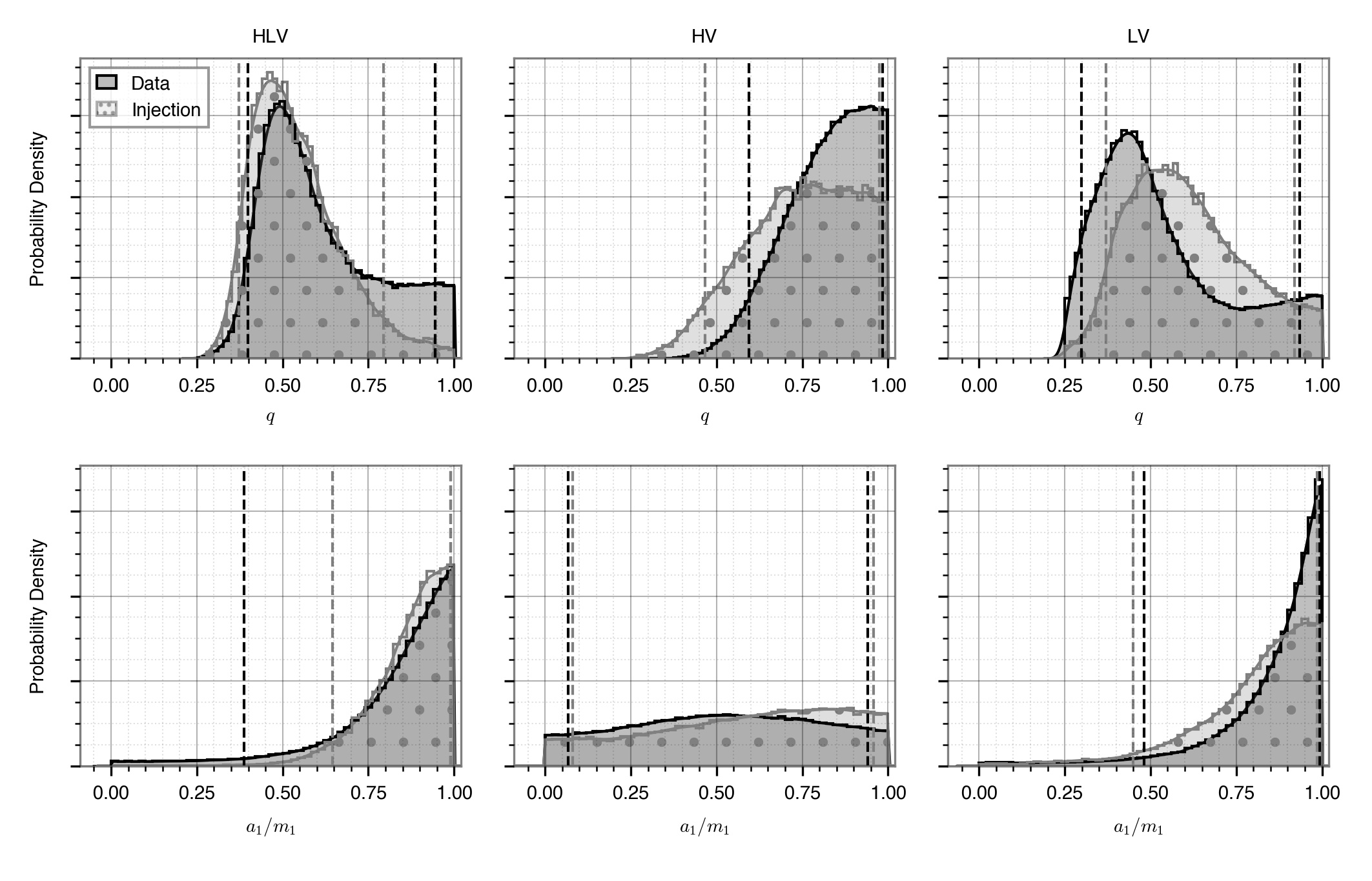}
  \caption{
  \label{fig:injection}
{\bf Comparison between GW200129 results and those from a model-waveform injection.} 
One-dimensional posterior distributions for the primary black hole's spin, $a_{1}/m_1$, and the binary's mass ratio, 
$q = m_2/m_1 \leq 1$, from a parameter-estimation analysis of GW200129 in raw detector data starting at 30\,Hz, and an idealised 
zero-noise injection. The main 
results for the three-detector network (left) are broadly consistent between the real data and the injection. We also find in both the 
real data and the injection that analysis of a Hanford-Virgo-only analysis (middle) prefers equal-mass binaries and much weaker
evidence for precession, while a Livingston-Virgo-only analysis (right) identifies an unequal-mass precessing binary; this can be
explained by the lower SNR in Hanford (\HSNRgwtc, vs \LSNRgwtc\, in Livingston), which reduces the measurability of precession.
  }
\end{center}
\end{figure}

To investigate these effects, we checked whether we would find similar results if the precessing-binary signal matching our preferred
parameters were observed in a detector network with zero noise. 
We performed a parameter-estimation run on an idealised example of a theoretical signal from a binary with our 
best-estimate parameters (those with the maximum likelihood in our main parameter-estimation results), as it would be observed in 
a network of detectors with the same frequency-dependent sensitivity, but no noise. For this exercise we compared with an analysis
on raw data, and starting at 30\,Hz. The results are shown in Extended Fig.~\ref{fig:injection}. We recovered parameters 
consistent with our measurements in real data: the spin-magnitude again ``rails'' against extreme spins and the mass-ratio measurement
has a tail that extends towards equal-mass systems. In addition, we again find that precession is identified in the 
 Livingston and Virgo detectors, but \emph{not} when repeated with only the Hanford and Virgo detectors. 
 This can be explained by the lower SNR in the Hanford detector: the precession SNR
will be only $\sim$2.4 in that detector, and so difficult to distinguish from noise. 

(3) Prior effects. The choice of priors for the spin magnitudes and tilt angles can strongly affect parameter estimates (see e.g. Ref.~\cite{Zevin:2020gxf}). 
Given the relatively small number of GW observations to date, there is no clear motivation to use a prior based on the currently observed population or
motivated by other astrophysical considerations. Indeed,
the LVK collaborations still maintain a prior that remains agnostic about black-hole masses and spin magnitudes and orientations~\cite{LIGOScientific:2021djp}. 
In our analysis we used the same priors as those used by the LVK collaborations, i.e., flat priors in the component masses and spin magnitude and the cosine of the 
misalignment angle. 
Of course, alternative spin priors consistent with different astrophysical binary formation scenarios can be
used to perform model selection on those scenarios (see e.g.~\cite{Farr:2017uvj, Tiwari:2018qch, Hoy:2021rfv}), and this would be an interesting topic for future work. 

(4) Noise effects. The LVK analysis noted some noise artifacts near the time of the observation, but mitigation procedures were applied to the 
data, and these de-glitched data were used for the LVK analysis. We have analysed both the raw and de-glitched data, and find broadly 
consistent results. However, the support for precession is slightly \emph{higher} in the raw data, and an interesting topic for future work
would be to quantify more precisely the effect of the de-glitching procedures on the precession measurement. 
As a final consistency check, we also perform injections of the preferred waveform into data within 2\,s of GW200129, and recovery of these 
signals also gives consistent results.


{\bf Data Availability.} The posterior samples from the analyses performed in this work are available on Zenodo: \href{https://zenodo.org/record/6672460}{https://zenodo.org/record/6672460}. Public documentation is available at the following URL: \href{https://data.cardiffgravity.org/GW200129-precession/}{https://data.cardiffgravity.org/GW200129-precession/}.

{ \ } 

{\bf Acknowledgements} 
We thank Thomas Dent, Shrobana Ghosh, Eleanor Hamilton, Panagiota Kolitsidou, Lionel London and Frank Ohme for useful discussions. 
We also thank Keith Riles for his guidance during the internal LIGO review process. The authors were supported in part by Science 
and Technology Facilities Council (STFC) grant ST/V00154X/1 and 
European Research Council (ERC) Consolidator Grant 647839.
Calculations were performed using the supercomputing facilities at Cardiff University operated by Advanced Research Computing 
at Cardiff (ARCCA) on behalf of the Cardiff Supercomputing Facility and the HPC Wales and Supercomputing Wales (SCW) projects. 
We acknowledge the support of the latter, which is part-funded by the European Regional Development Fund (ERDF) via the Welsh 
Government. In part the computational resources at Cardiff University were also supported by STFC grant ST/I006285/1.
We are also grateful for computational resources provided by LIGO Laboratory and supported by National Science Foundation Grants 
PHY-0757058 and PHY-0823459.
This material is based upon work supported by NSF's LIGO Laboratory, which is a major facility fully funded by the National Science 
Foundation.
This research has made use of data, software and/or web tools obtained from the Gravitational Wave Open Science Center 
(\href{https://www.gw-openscience.org}{https://www.gw-openscience.org}), a service of LIGO Laboratory, the LIGO Scientific Collaboration 
and the Virgo Collaboration. LIGO is funded by the U.S. National Science Foundation. Virgo is funded by the French Centre National de 
Recherche Scientifique (CNRS), the Italian Istituto Nazionale della Fisica Nucleare (INFN) and the Dutch Nikhef, 
with contributions by Polish and Hungarian institutes.

Plots were prepared with Matplotlib \citep{2007CSE.....9...90H}, GWpy \citep{gwpy} and {\sc{PESummary}} ~\citep{Hoy:2020vys}. 
Parameter estimation was performed with the {\sc{LALInference}} \citep{Veitch:2014wba} and {\sc{LALSimulation}} libraries
within {\sc{LALSuite}} \citep{lalsuite}, as well as
the BILBY \citep{Ashton:2018jfp, Romero-Shaw:2020owr} and PBILBY Libraries \citep{Smith:2019ucc} and the DYNESTY nested sampling package \citep{Speagle:2019ivv}. {\sc{NumPy}}~\cite{numpy}, {\sc{Scipy}}~\cite{mckinney-proc-scipy-2010} and {\sc{positive}}~\cite{London:2018nxs, lionel_london_2020_3901856} were used during the analysis.

{ \ }

\noindent {\bf Author contributions} 

\noindent M.H. initiated and lead the study and writing of the text. C.H. and J.T. performed the parameter estimation calculations on the detector data, and the synthetic injections. J.T. performed the mismatch calculations. V.R. and
S.F. provided valuable input to the analysis; V.R. verified the PE results and S.F. produced Fig. 3. M.H., C.H., J.T., S.F. and V.R. all contributed
to the interpretation of the results and writing the paper. 
Although they did not directly contribute to the results presented here, the other authors contributed to the LVK's initial understanding of GW200129, 
which was the foundation of this work.

{ \ }

\noindent {\bf Competing interests}

\noindent The authors declare that they have no competing financial interests.

\end{document}